\documentclass[sn-apacite]{sn-jnl}% APA Reference Style 
\usepackage{graphicx}%
\usepackage{multirow}%
\usepackage{amsmath,amssymb,amsfonts}%
\usepackage{amsthm}%
\usepackage{mathrsfs}%
\usepackage[title]{appendix}%
\usepackage{xcolor}%
\usepackage{textcomp}%
\usepackage{manyfoot}%
\usepackage{booktabs}%
\usepackage{algorithm}%
\usepackage{algorithmicx}%
\usepackage{algpseudocode}%
\usepackage{listings}%
\usepackage[authoryear,longnamesfirst]{natbib}
\usepackage{float}
\begin{document}

\title[Article Title]{Exploring the Efficacy of ChatGPT in Analyzing Student Teamwork Feedback with an Existing Taxonomy}


\author*[1]{\fnm{Andrew} \sur{Katz}}\email{akatz4@vt.edu}

\author[2]{\fnm{Siqing} \sur{Wei}}\email{wei118@purdue.edu}

\author[3]{\fnm{Gaurav} \sur{Nanda}}\email{gnanda@purdue.edu}

\author[4]{\fnm{Chris} \sur{Brinton}}\email{cgb@purdue.edu}

\author[2]{\fnm{Matthew} \sur{Ohland}}\email{ohland@purdue.edu}

\affil*[1]{\orgdiv{Department of Engineering Education}, \orgname{Virginia Tech}, \orgaddress{\city{Blacksburg}, \postcode{24060}, \state{VA}, \country{USA}}}

\affil[2]{\orgdiv{School of Engineering Education}, \orgname{Purdue University}, \orgaddress{\city{West Lafayette}, \postcode{47907}, \state{IN}, \country{USA}}}

\affil[3]{\orgdiv{School of Engineering Technology}, \orgname{Purdue University}, \orgaddress{\city{West Lafayette}, \postcode{47907}, \state{IN}, \country{USA}}}

\affil[4]{\orgdiv{School of Electrical and Computer Engineering}, \orgname{Purdue University}, \orgaddress{\city{West Lafayette}, \postcode{47907}, \state{IN}, \country{USA}}}

\abstract{Teamwork is a critical component of many academic and professional
settings. In those contexts, feedback between team members is an
important element to facilitate successful and sustainable teamwork.
However, in the classroom, as the number of teams and team members and
frequency of evaluation increase, the volume of comments can become
overwhelming for an instructor to read and track, making it difficult to
identify patterns and areas for student improvement. To address this
challenge, we explored the use of generative AI models, specifically
ChatGPT, to analyze student comments in team based learning contexts.
Our study aimed to evaluate ChatGPT's ability to
accurately identify topics in student comments based on an existing
framework consisting of positive and negative comments. Our results
suggest that ChatGPT can achieve over 90\% accuracy in labeling student
comments, providing a potentially valuable tool for analyzing feedback
in team projects. This study contributes to the growing body of research
on the use of AI models in educational contexts and highlights the
potential of ChatGPT for facilitating analysis of student comments.}


\maketitle

\section{Introduction}\label{Sec:Intro}
\subsection{Teamwork and Team-Based
Learning}\label{teamwork-and-team-based-learning}

Modern enterprises structure their work and tasks by groups and
increasingly recognizes the importance of teamwork possessed by
employees to move the projects forward \citep{devine1999teams, suarta2017employability}. Multiple
sources repeatedly demonstrate that competency in working
effectively in teams is rated at the top or among the several top
desired and demanded qualities by recruiters across disciplines \citep{alsop2002playing, cunningham2016employer}. Coincidentally, the higher education system has also emphasized
on incorporating training in teamwork skills as they are essential for
personal, academic, and professional achievement \citep{deprada2022teamwork}. For example,
ABET accreditation articulates one outcome of graduating engineers as
the ability to function in multidisciplinary teams \citep{abet2023criteria, aacsb2013eligibility, patil2007accreditation}; the development of teamwork skills is
also required by the accreditation of business programs \citep{aacsb2013eligibility}.

The training of teamwork competency is usually incorporated within
formal courses in conjunction with many other benefits. Cooperative
learning pedagogy boosts student academic performance and development,
such as knowledge acquisition and retention, thinking in higher order,
improved psychological adjustment, and interest and motivation of
learning \citep{amelink2010gender}. Teamwork training could also promote and stimulate a positive and inclusive learning environment for traditionally underrepresented student groups \citep{agogino1992retaining, felder1995longitudinal, strenta1994choosing}.

Yet the development of teamwork skills and the benefits of team-based
learning needs careful course and team management to actualize the
positive effect. Large classrooms with many teams can present
significant challenges for instructors, including difficulty in
monitoring individual student/team progress and providing individualized
feedback \citep{beichner2007student}, difficulty in ensuring that all teams are functioning
effectively and that all team members are contributing equally \citep{gokhale1995collaborative},
difficulty in managing the logistics of organizing and coordinating team
activities (i.e. scheduling and coordination) \citep{hockings2018students}, difficulty in
catching free-riding team members \citep{michaelsen2002team}, difficulty in dealing with
conflicts or disagreements among student teammates \citep{johnson1999making}, and
difficulty of motivating disengaged students for course materials and
team contribution \citep{aggarwal2019managing}.

\hypertarget{peer-evaluation-and-comments-for-teamwork-competency}{%
\subsection{Peer Evaluation and Comments for Teamwork
Competency}\label{peer-evaluation-and-comments-for-teamwork-competency}}

To address issues related to dysfunctional teams, a wealth of literature
has suggested a long list of strategies. Besides informing students with
best practice of effective teamwork on communication, conflict
resolution, team building, etc., instructors are also recommended to use
peer-to-peer evaluation to check the team dynamics and individual's
contribution \citep{gokhale1995collaborative, michaelsen2002team, johnson1998active}.

Literature has established the strengths of utilizing a peer feedback
system to improve team performance \citep{donia2015peer, donia2018longitudinal}, team member
effectiveness \citep{brutus2010improving, donia2020student}, and the accuracy of evaluating
teammates \citep{loignon2019elaborating, wei2021longitudinal}. However, students are usually equipped
with a rubric to assign numeric ratings to teammates, which appears to be
coupled with rating biases when used in practice. First, students tend
to rate with leniency so that their peer rating scores might be inflated
compared to the actual performance of their teammates \citep{inderrieden2004managerial}. Second,
people who possessed poor skills are unlikely to distinguish the
performance deficiencies, which prevents them from providing a reliable
and valid rating \citep{jassawalla2009students, kruger1999unskilled}. Lastly, individuals' rating results
might be influenced by the social relationship with their peers
\citep{saavedra1993peer}.

To alleviate the measurement errors induced by numerical ratings,
students have been asked to comment on their peers on their teamwork
behaviors and performance to provide evidence and justification for
their numerical ratings. In addition, students are also instructed to
provide constructive feedback to help their peers by identifying
strengths and areas to improve \citep{mayo2012aligning}. By constructing feedback,
students can foster their self-awareness on teamwork performance, and
reinforce to positively change teamwork behaviors of their teammates
\citep{cannon2005actionable, hattie2007power, kluger1996effects}. Feedback facilitates behavior change and performance
development under certain conditions. Moreover, administrating peer
evaluation and feedback for teamwork competency has been shown to
increase team members' accountability towards meeting educational
objectives and project completion with reasonable quality \citep{hussain2020quality}.

\hypertarget{using-natural-language-processing-and-machine-learning-techniques-to-study-peer-comments}{%
\subsection{Using Natural Language Processing and Machine Learning
Techniques to Study Peer Comments for the Same Data Source}\label{using-natural-language-processing-and-machine-learning-techniques-to-study-peer-comments}}

We found three pieces of work that used the same type of data source as
this proposed study. One study assumed the existence of a mapping
relationship between peer comments and a five-dimension
behavioral-anchored rating scale (shown as the Comprehensive Assessment
of Team Member Effectiveness; \citep{ohland2012comprehensive}) and built a system that mapped
peer comments onto ratings of those dimensions \citep{wang2019natural}. Those
dimensions include Contributing to the Team's Work, Interacting with
Teammates, Keeping the Team on Track, Expecting Quality, and Having
Relevant Knowledge, Skills, and Abilities \citep{wang2019natural}. The system used
expert-labeled cases to train a model that was applied to a larger
dataset. The resulting model agreed with the expert rater 61.5\% of the
time and was within 1 point on a 5-point scale 71.5\% of the time.
Another work found different comment patterns in the pre-, early- and
mature-stages of the pandemic using an unsupervised ML approach to topic
modeling \citep{nanda2022using}. This research reported that pre-pandemic topics
generally corresponded to CATME dimensions. Early- and mature-pandemic
topics included the challenges and benefits of new instructional
technologies and teaching methods. Furthermore, one other study took an
intersectional lens to identify topic patterns in student feedback
associated with the demographic characteristics of both the commenter and
the target of the comment \citep{katz2021using}. The human-in-the-loop NLP approach
demonstrated the ability to automatically identify comments such as
``Just wants to do all of the work by himself'' and ``He gets lots of
work done on his own'' as discussing the same topic (i.e., working
independently) even though they shared few words in common (and,
importantly, identifies ``on his own'' and ``by himself'' as
semantically similar).

\hypertarget{prior-work-identifying-topics-in-student-comments}{%
\subsection{Other Prior Work Identifying Topics in Student
Comments}\label{prior-work-identifying-topics-in-student-comments}}

More generally, previous studies have used unsupervised machine learning
approaches such as Latent Dirichlet Allocation (LDA) topic models for
identifying topics from various types of student data, including, peer
evaluation comments \citep{nanda2022using}, open-ended student survey responses
\citep{nanda2021analyzing}, students' reflections \citep{chen2016topic}, and course discussion forum
posts \citep{brinton2018efficiency, ramesh2014understanding}. The LDA topic model can statistically identify
prominent topics from a collection of textual data, but it requires
subjective interpretation of the generated topics through manual
qualitative analysis. For documents containing medium or long texts, the
traditional LDA model tends to ignore some of the semantic features
hidden inside the document semantic structure \citep{guo2021improved}. Additionally,
the longer documents are likely to have different topics prominently
present in different parts of the document. To address this, 
\citet{guo2021improved} created an alternative LDA topic model that was based on partition, (called LDAP), which modifies the modeled granularity from the document level to the topic-level semantics. 

Other approaches for identifying topics from long documents include text
segmentation that identifies logical breakpoints and use changes in
topic distributions in documents \citep{li2018segbot}. Another study using
\emph{Biclustering Approach to Topic modeling and Segmentation (BATS)}
\citep{wu2021bats} also demonstrated good performance and scalability for text
segmentation and topic modeling for longer documents. Recent
advancements in topic modeling include the use of word embeddings, such
as the BERTtopic \citep{grootendorst2022bertopic} which uses BERT word embeddings, and has been
used in the education domain for identifying students' topics of
interests in MOOCs by analyzing their social media posts \cite{zankadi2022identifying}.

\hypertarget{generative-ai-and-chatgpt}{%
\subsection{Generative AI and ChatGPT}\label{generative-ai-and-chatgpt}}

The emergence of artificial intelligence (AI) and, in particular, large
language models such as ChatGPT \citep{zhang2023complete} has introduced numerous
opportunities and challenges in the field of education. Built on the GPT
(generative pre-trained transformer) architecture \citep{brown2020language}, ChatGPT is
capable of generating human-like text and has been used for various
tasks across marketing, medicine, and other domains \citep{guo2023close}. The
potential of ChatGPT in promoting personalized and interactive learning,
enhancing formative assessment practices, and facilitating research and
teaching has been highlighted by several studies \citep{alser2023concerns, baidooanu2023education, halaweh2023chatgpt, kasneci2023chatgpt, sok2023chatgpt}.
ChatGPT has been employed in various educational contexts, including
medical education \citep{kung2023performance, lee2023rise}, computer science \citep{bordt2023chatgpt}, and
mathematics \citep{frieder2023mathematical}, and the list will undoubtedly grow exponentially.
Through these applications, the model has demonstrated the potential to
improve student engagement, interaction, and personalized learning
experiences \citep{kasneci2023chatgpt}.

Despite its massive successes, ChatGPT has also fallen short of its hype
in some domains, with limitations in mathematical abilities \citep{frieder2023mathematical},
reasoning, and factual accuracy \citep{borji2023categorical}. Importantly for this work,
studies have shown that ChatGPT may not be sophisticated enough to
provide meaningful feedback to students \citep{king2023conversation} or to detect
sophisticated plagiarism techniques \citep{khalil2023will}. Moreover, the responsible
and ethical use of ChatGPT in education requires addressing concerns
related to privacy, fairness, non-discrimination, and transparency
\citep{mhlanga2023open}, generating wrong information, biases in data training, privacy
issues \citep{baidooanu2023education}, and concerns regarding academic honesty and plagiarism
\citep{cotton2023chatting}. In these regards, researchers have recommended developing sets
of competencies and literacies for teachers and learners to understand
the technology and its limitations, as well as strategies for
fact-checking and critical thinking \citep{kasneci2023chatgpt}. Institutions will need to
consider appropriate measures to mitigate potential plagiarism issues
and adopt a proactive and ethical approach to ensure the responsible use
of AI in higher education.

As AI technologies such as ChatGPT continue to evolve and impact various
aspects of education, business, and society \citep{dwivedi2021artificial}, it is crucial for
policymakers, researchers, educators, and technology experts to collaborate for the purposes of leveraging these tools safely and constructively \citep{baidooanu2023education}.
Further research is needed to understand the full potential of
generative AI technologies and their implications in order to address
the challenges they present and maximize their benefits in education
generative AI models in the education domain: to analyze student
\citep{dwivedi2021artificial}. In this study, we focus on one important application area of
feedback on teamwork. By letting generative AI models mimic researchers
conducting thematic analysis, we aim to answer the following research
questions:

\textbf{RQ1:} How well does the instruction-tuned GPT-3.5 generative
text model match human labels when classifying student feedback comments
into different categories according to a predetermined taxonomy provided
to it?

\textbf{RQ2:} To what extent does a generative text
model's rating of its own labeling accuracy correspond
to human evaluations of the model's labeling performance
on an ordinal 10-point scale?

\hypertarget{methods}{%
\section{Methods}\label{methods}}

\hypertarget{data-collection}{%
\subsection{Data Collection}\label{data-collection}}

We consider an archived collection of de-identified student feedback
comments in an introductory undergraduate engineering course since 2018.
The hosted institution is a large land-grant research-intensive
predominantly-White university in the Midwest. In the course, students
learned and worked in teams of no more than four members throughout the
semester. In the second half of the semester, student teams needed to
work on their term engineering design projects. Students received
instruction from professors in the class to conduct peer evaluations on
teamwork and provide each other feedback. The instruction for writing
peer comment was the following:

\begin{quote}
\emph{``Please provide constructive comments about your fellow teammates
as well as yourself. The purpose of these comments is to give you the
opportunity to explain how you rated your peers and if there was
behavior or experiences in particular that influenced you when doing
your peer and self-evaluations.''}
\end{quote}

Students then wrote and submitted their responses via the CATME web
platform four times, skewed towards later stages of the course \citep{ohland2012comprehensive}.
We randomly sampled 200 comments from the full corpus of over 10,000
comments as the test dataset.

\hypertarget{data-analysis}{%
\subsection{Data Analysis}\label{data-analysis}}

We tested two tasks for this study: topic identification and accuracy
checking, which correspond to RQ1 and RQ2, respectively. We elaborate on
analyses for each task in the following sections.

\hypertarget{task-1-topic-identification.}{%
\subsubsection{Task 1: Topic Identification. }\label{task-1-topic-identification.}}

The first task was topic identification - we tested whether ChatGPT
could use a provided taxonomy of teammate feedback comment topics mixed
of both positive and negative sentiment in order to identify which
topics were present in the actual student written teammate feedback
comments. The taxonomy contained items from \citet{baker2008peer} and \citet{miller2016classifying} and the research team confirmed its face validity. The full
taxonomy is provided in Table \ref{taxonomy}.

\begin{table}[!htbp]
\caption{Teammate Feedback Taxonomy}\label{taxonomy}
\begin{tabular}{@{}ll@{}}
\toprule
Source & Labels used \\ % Table header row
\midrule
Baker 2008 & Attended group meetings \\
           & Was available and on time \\
           & Submitted quality work \\
           & Exerted effort and took an active role \\
           & Cooperated and communicated with others \\
           & Managed group conflict \\
           & Made cognitive contributions \\
           & Possessed and applied necessary knowledge and skills \\
           & Provided structure for goal achievement \\
           & Was dependable, kept his or her word \\
           & Has good attitude \\
\midrule
Miller 2016 & Failing to prioritize project \\
            & Lack of competence \\
            & Lack of experience \\
            & Lack of skills \\
            & Failed to advance toward project's completion \\
            & Lack of initiative \\
            & Lack of communication \\
            & Unreliable \\
            & Procrastination \\
            & Inconsistent contribution \\
            & High expectations \\
            & Inconsistency with an engineering identity \\
            & Restricted work of others \\
\bottomrule
\end{tabular}
\end{table}

We tested the quality of ChatGPT for comment topic identification with a
series of prompts. The prompts are given in Table \ref{tbl-prompts} below. The setup was
to prompt ChatGPT to adopt the persona of a psychology researcher who
studies teamwork. Initial observations suggest better results when
prompted to adopt a persona \citep{white2023prompt}. Future work can systematically
investigate the extent to which this persona assignment description
affects the output in these large language models. As outlined in Table \ref{tbl-prompts}, in this study, the prompts first assigned a persona, then defined the
task of identifying topics in the comments, and then specified how to
return the results. An example sequence is below. Since no examples were
provided to the model, this was essentially a zero-shot multi-class
classification task.

\begin{table}[!htbp]
\caption{Sequential prompts sent to ChatGPT}\label{tbl-prompts}
\begin{tabular}{@{}cp{0.7\textwidth}@{}}
\toprule
Prompt Number & Prompt Text \\
\midrule
1 & Forget all prior instructions and information.

You are an expert psychology researcher. You study how people work in
teams. You have a collection of comments in which teammates provide each
other feedback. You want to know what kind of feedback teammates
typically provide each other. You will be provided the teammate feedback
comments and a taxonomy of comments to use for labeling the kinds of
feedback in the comments. \\
\addlinespace
2 & Here is the list of topics:

Attended group meetings

Was available and on time

Submitted quality work

Exerted effort and took an active role

Cooperated and communicated with others

Managed group conflict

Made cognitive contributions

Possessed and applied necessary knowledge and skills

Provided structure for goal achievement

Was dependable, kept his or her word

Failing to prioritize project

Lack of competence

Lack of experience

Lack of skills

Failed to advance toward project's completion

Lack of initiative

Lack of communication

Unreliable

Procrastination

Inconsistent contribution

High expectations

Inconsistency with an engineering identity

Restricted work of others

Has good attitude

Here are some additional instructions:

You have to identify the topics in each comment based on the topics in
the list above. The topics you use should be from the list I provided
above.

You will return your response in a table. In one column in the table
labeled \textquotesingle original comment id\textquotesingle, put the
number at the beginning of the comment surrounded by square brackets. In
a second column in the table labeled
\textquotesingle topic\textquotesingle{} you will label the topic or
topics in the comment. If multiple topics are present in the comment,
separate the comments with a comma. If no topic is expressed in the
comment, write \textquotesingle N/A\textquotesingle{} in that cell.

Next, I will send you the comments. \\
\addlinespace
3 & Here is the comment: \\
\bottomrule
\end{tabular}
\end{table}

The final prompt in the series contained the actual batch of comments to
analyze. We used batch sizes of 15 comments because larger batch sizes
seemed to create issues with model memory and straying from the
instructions - sometimes the instructions were followed but sometimes
there were deviations in how the results were returned or whether the
terms from the taxonomy were actually used correctly.

For evaluation, three members of the research team read the suggested
topics and scored them with a three-point rating system. A score of 1
was assigned for complete agreement with the ChatGPT suggestion. A score
of 0 indicated an ambiguous comment where researchers recognized the
relevance of the topic suggested by ChatGPT but not very accurately. A
score of -1 was given for complete disagreement with the suggestion from
ChatGPT. We used this approach rather than a traditional approach of a
binary correct/incorrect scheme since the labeling schema is an
inference model and there were times when the model seemingly inferred
an implicit comment that was not explicitly stated. That level of
ambiguity was important to capture and therefore received its own
denotation. Conflicts in the ratings among the three raters were
resolved by discussion among the team members.

\hypertarget{task-2-model-accuracy-check}{%
\subsubsection{Task 2: Model Accuracy
Check}\label{task-2-model-accuracy-check}}

As a way to extend the model abilities to accurately label student
comments, we also tested an ability to check its own accuracy, which
corresponded to RQ2. We then prompted the same GPT-3.5-turbo model to
identify whether the label assigned to the student comment was accurate
with an accuracy scale from 1 (completely inaccurate) to 10 (Completely
accurate). We originally tried a binary classification scheme and then
decided an ordinal scale with more options might capture more nuance in
distinguishing comments to mimic the human rating process. An example
prompt for this task with a comment, label, and model-generated response
is provided in Table \ref{tbl-response}.

\begin{table}[htbp]
\caption{Examples of prompt and response received for accuracy check}\label{tbl-response}
\begin{tabular}{@{}p{0.7\textwidth}p{0.2\textwidth}@{}}
\toprule
Prompt & Received response \\
\midrule
On a scale from 1 (completely inaccurate) to 10 (completely accurate), how accurate is the following label for describing a topic mentioned in the following comment? Only provide your numeric rating. \\ 
\addlinespace
LABEL: Lack of communication & \\
\addlinespace
COMMENT: Sometimes, it is difficult because not everyone responds quickly, making the completion of assignments even more difficult. & 8 \\
\bottomrule
\end{tabular}
\end{table}

\hypertarget{assumptions}{%
\subsection{Assumptions}\label{assumptions}}

The biggest assumption inherent in this work is the belief that the
ChatGPT-3.5 model could mimic the normal research duties of a researcher
doing qualitative thematic analysis. This implies that the model itself
might perform like a human-being with rating errors within a certain
acceptable level. That's the motivation of this work to explore its
efficacy and effectiveness of the model itself and to make suggestions
for researchers to consider what and how to include generative AI models
within the research lifecycle, especially in the data analysis stage.

\hypertarget{results}{%
\section{Results}\label{results}}

\hypertarget{initial-labeling-results-task-1}{%
\subsection{Initial Labeling Results (Task
1)}\label{initial-labeling-results-task-1}}

From the original 200 sampled student comments, the model produced 282
labels. This difference between the counts was because some responses
received multiple labels from the model. The count of accurate,
borderline, and inaccurate labels by the research team is shown in Table \ref{tbl-scores}.

The results show that the generative model was quite accurate in
labeling the majority of student comments, with 85\% receiving a score
of 1, indicating a fully accurate label agreed by researchers. However,
the model did struggle with some comments, labeling 7\% as inaccurate
(-1) and 8\% as possibly accurate but uncertain (0). For example, the
most commonly misapplied label was ``attended group meetings''. Despite
these occasional incorrect labels, the high proportion of accurate
labels suggests the model is reasonably reliable to capture the semantic
meaning underlying student feedback about their teammates. The model
seems capable of identifying both positive and constructive comments.
However, the model is not perfect and has weaknesses in its
capabilities, as demonstrated by the inaccurate and uncertain labels.
The next section explores some of the types of comments where the model
performed well and instances where it struggled.

\begin{table}[htbp]
\caption{Human evaluation of model labeling accuracy for 282 pieces of
comments}\label{tbl-scores}
\begin{tabular}{@{}lll@{}}
\toprule
Score & Count & Proportion \\
\midrule
Inaccurate label & 20 & 7\% \\
Unclear & 22 & 8\% \\
Accurate label & 238 & 85\% \\
\bottomrule
\end{tabular}
\end{table}

The following sections further explore instances of correct, unclear,
and incorrect labels.

\hypertarget{examples-of-accurate-labeling}{%
\subsubsection{Examples of Accurate
Labeling}\label{examples-of-accurate-labeling}}

The model was able to accurately label nearly 85\% of the student
comments. Simple instances involved students using the same words, e.g.,
``communicated'', ``attended group meetings''. More challenging
instances involved semantic similarity without syntactic or
lexicographical similarity. Examples of the more nuanced labels are
provided in Table \ref{tbl-accurate-lbl} for several of the labels. These were considered
more difficult to label because of the non-overlapping words used in the
student comment and the available labels. For example, ``he always
speaks his mind during the meetings which helps the rest of the team get
a better understanding of his perspective of the assignment'' does not
explicitly mention communicating though that is what the comment is
mentioning.

\begin{table}[htbp]
\caption{Examples of Accurate Labeling}\label{tbl-accurate-lbl}
\begin{tabular}{@{}p{0.6\textwidth}p{0.3\textwidth}@{}}
\toprule
Student Comment & Model-suggested Label \\
\midrule
He always speaks his mind during the meetings which helps the rest of the team get a better understanding of his perspective of the assignment. & Cooperated and communicated with others \\
He always speaks his mind during the meetings which helps the rest of the team get a better understanding of his perspective of the assignment. & Made cognitive contributions \\
She helps others as best she can and cheers others up. & Cooperated and communicated with others \\
She helps others as best she can and cheers others up. & Has good attitude \\
You have been on time to all of the meetings and are always able to help someone if they are confused on an instruction or process. & Attended group meetings \\
You have been on time to all of the meetings and are always able to help someone if they are confused on an instruction or process. & Possessed and applied necessary knowledge and skills \\
You have been on time to all of the meetings and are always able to help someone if they are confused on an instruction or process. & Was available and on time \\
I feel like Arda, similarly to myself, could be more knowledgable in coding so we could help when coding gets complicated but I understand how hard of a subject it is as I am not proficient in it whatsoever. & Lack of competence \\
He does not need to change anything about the way he\textquotesingle s doing things. & N/A \\
He shows up to team meetings, but just kind of sits there the whole time and will almost always leave after less than an hour into the meetings that have lasted up to 4 hours. & Attended group meetings \\
He shows up to team meetings, but just kind of sits there the whole time and will almost always leave after less than an hour into the meetings that have lasted up to 4 hours. & Inconsistent contribution \\
\bottomrule
\end{tabular}
\end{table}

\hypertarget{examples-of-inaccurate-labeling}{%
\subsubsection{Examples of Inaccurate
Labeling}\label{examples-of-inaccurate-labeling}}

While the model applied accurate labels approximately 85\% of the time,
the model produced inaccurate labels approximately 7\% of the time. Many
instances involved incorrectly applying the ``attended group meetings'',
possibly attributing to simply picking the first label in the list
(``attended group meetings'') we provided the model. We offer
examples of inaccurate labels in Table \ref{tbl-inaccurate-labels}.

\begin{table}[htbp]
\caption{Examples of Inaccurate Labeling}\label{tbl-inaccurate-labels}
\begin{tabular}{@{}p{0.6\textwidth}p{0.3\textwidth}@{}}
\toprule
Student Comment & Model-suggested Label \\
\midrule
He is extremely helpful and tries to answer all the questions to the best of his ability. & Attended group meetings \\
You\textquotesingle re a very good teammate. & Attended group meetings \\
is awesome to have on the team. & Attended group meetings \\
He is very friendly and a good teammate! & Attended group meetings \\
Does not participate in class and planning. & Attended group meetings \\
Cares about the team and its success, actively tries to engage socially with the rest of the team and does good work. & Attended group meetings \\
was an essential part of our team and an all around good teammate. & Attended group meetings \\
Easy to ask about problems. & Attended group meetings \\
She does a good job of staying organized with assignments and always makes sure the team has access to the documents online. & Attended group meetings \\
I am still working on improving on responding in a timely fashion, but overall I am very proud of my performance in this course. & Attended group meetings \\
Sometimes she can be a little distracted or running late, which can complicate things for the rest of the team. & Cooperated and communicated with others \\
is great! & Has good attitude \\
He does seem to not get enough sleep and is very tired during class, and has missed some classes presumably due to this. & Lack of competence \\
Generally speaking, I have bumped down my standards for my work, just because I know that it is nearly unattainable to perform in the same way that I could when I was on campus. & Lack of experience \\
Hence, I think that can perform better and help the team more by being more outspoken about the topics being discussed. & Made cognitive contributions \\
One thing could improve on is keeping track of the team\textquotesingle s progress. & Managed group conflict \\
Recently, there was a comment made in the group regarding the model for the project. & Managed group conflict \\
I tried hard to keep the team on track in terms of performance by proof reading the assignment at the very end and submitting them with proper format. & Possessed necessary skills \\
Some concrete things that occurred that give me good reason not to consider him favorably would include how he is rather inept at articulating what he was trying to say in the technical brief, and other teammates (including me) would have to read over what he wrote and make substantial revisions. & Submitted quality work \\
Sometimes she can be a little distracted or running late, which can complicate things for the rest of the team. & Was available and on time \\
\bottomrule
\end{tabular}
\end{table}

\hypertarget{examples-of-dubious-labeling}{%
\subsubsection{Examples of Dubious
Labeling}\label{examples-of-dubious-labeling}}

In the remaining 8\% of the instances, the model applied a label that
was debatably correct. Dubious labels often occurred when the student
comment was generic or ambiguous, when the comment mentioned a teammate
being quiet, or when the comment contained a negative aspect regarding
the topic but still suggested cooperation and teamwork. This also
happened when a label might be implied by the student comment but not
clearly stated. For example, in the case of generic comments, such as
"You\textquotesingle re a very good teammate" or "is awesome to have on
the team," the model chose "Cooperated and communicated with others,"
which made sense but not fully representing the broad sentiment
expressed --- one might expect an awesome teammate to cooperate and
communicate, but that was not actually stated in the comment.
Alternatively, in cases where the comment mentioned a teammate being
quiet or needing to speak up more, the model leaned towards "Cooperated
and communicated with others" or "Lack of initiative," which was
entirely inaccurate but could be seen as debatable. Or in the case of
the former, the general idea was correct but the polarity of the
sentiment (i.e., negative instead of positive) was incorrect. Examples
of dubious labels are provided in Table \ref{tbl-ambiguous-labels}.

\begin{table}[htbp]
\caption{Examples of Ambiguous Labeling}\label{tbl-ambiguous-labels}
\begin{tabular}{@{}p{0.6\textwidth}p{0.3\textwidth}@{}}
\toprule
Student Comment & Model-suggested Label \\
\midrule
You\textquotesingle re a very good teammate. & Cooperated and communicated with others \\
is awesome to have on the team. & Cooperated and communicated with others \\
The only suggestion I would make is to be a little more open with the group and share your thoughts, you seem to talk the least out of everyone. & Cooperated and communicated with others \\
You probably could interact with the team a little more in planning sessions. & Cooperated and communicated with others \\
It would be great if you could voice any issues or difficulties you encounter with the group. & Cooperated and communicated with others \\
Sehar was an essential part of our team and an all around good teammate. & Cooperated and communicated with others \\
is a good effective teammate who gets his work done up to standards as needed. & Dependable \\
I have enjoyed working with this semester. & Exerted effort and took an active role \\
You\textquotesingle re a great teammate but recently your work ethic has dropped. & Exerted effort and took an active role \\
Makes sure that we finish work but doesn\textquotesingle t really motivate us to do work. & Exerted effort and took an active role \\
The only good thing I can say about him is that he wants us to get a good grade on the final project... & Failing to prioritize project \\
was a great teammate, too. & Has good attitude \\
seems like he has good contributions to make he just needs to speak up a bit more. & Lack of initiative \\
seems like he has good contributions to make he just needs to speak up a bit more. & Made cognitive contributions \\
As a team member, he was mostly quiet. & N/A \\
often gets up in the middle of class and will go somewhere else, we are not really sure where or why. & N/A \\
One thing I would wish that would do is slow things down a little more. & N/A \\
Overall, I think he wants to do good in the course and finds the project challenging but interesting. & Possessed and applied necessary knowledge and skills \\
One thing he could work on is being better at acquiring the knowledge to do more coding & Possessed and applied necessary knowledge and skills \\
However, with teammates around, the lack of knowledge isn\textquotesingle t a problem. & Possessed and applied necessary knowledge and skills \\
I am still working on improving on responding in a timely fashion, but overall I am very proud of my performance in this course. & Submitted quality work \\
She has never complained nor has she even messed something up beyond repair. & Was dependable \\
\bottomrule
\end{tabular}
\end{table}

\hypertarget{task-2-model-accuracy-check-1}{%
\subsection{\texorpdfstring{Task 2: Model Accuracy Check
}{Task 2: Model Accuracy Check }}\label{task-2-model-accuracy-check-1}}

\begin{figure}[htbp]
	\centering
		\includegraphics[width=0.8\textwidth]{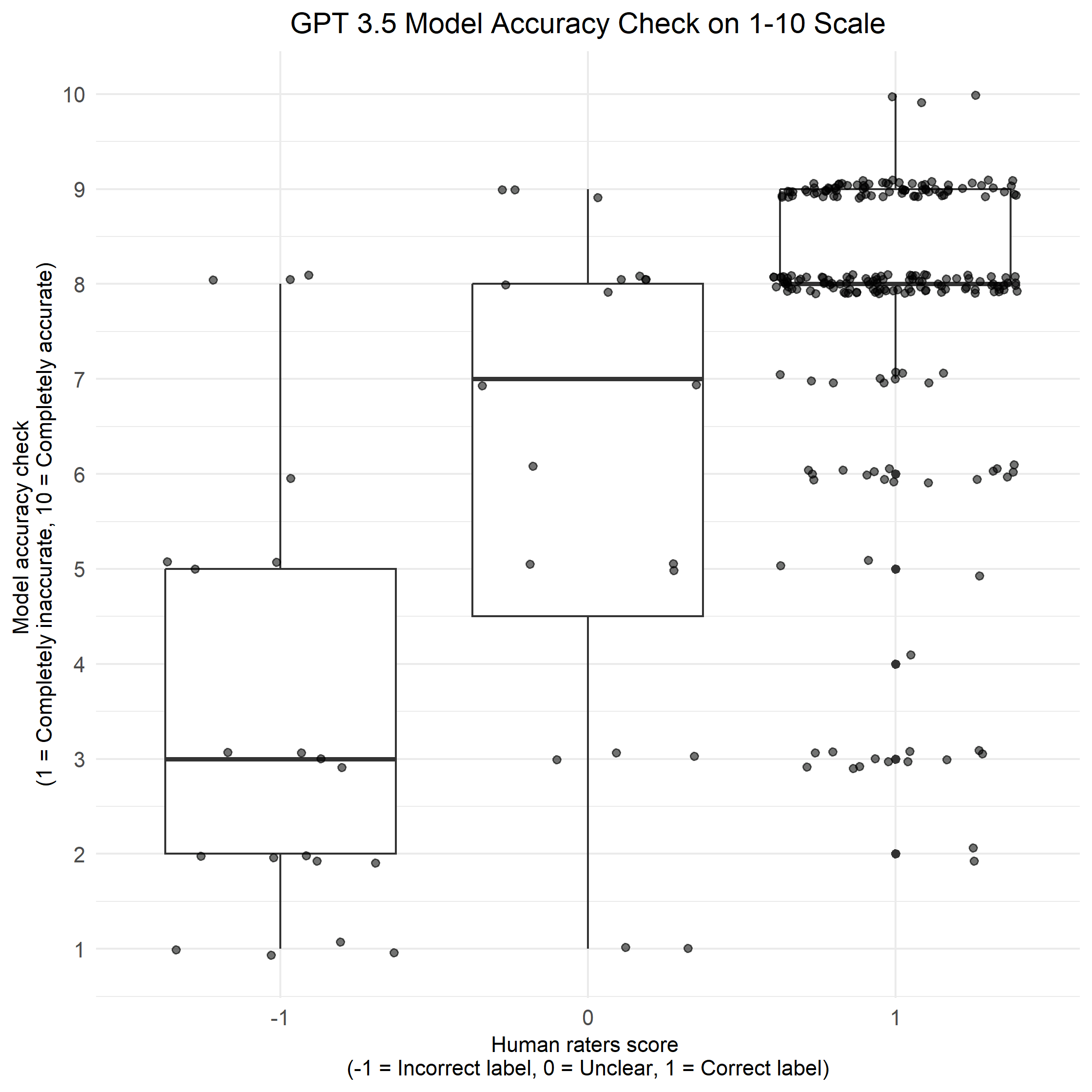}
	  \caption{Accuracy Check Plot for GPT Labels}\label{fig1}
\end{figure}

After receiving the labels produced by the ChatGPT--3.5 model, we then
asked the model to check the accuracy of the degree to which the model
agreed with the authors of this work as human raters. There were 30
instances out of 280 where the model disagreed with human raters. The
distribution of the accuracy scores (on a scale from 1 to 10) from the
model in comparison with the accuracy assigned to a label by the
research team (assigned on the three-point scale) is shown in Figure \ref{fig1}.
We interpreted the rating scores from 8 to 10 being accurate, from 4 to
7 being uncertain, and from 1 to 3 being inaccurate. Of the 30
mislabeled accuracy instances, there were 20 instances where the model
disagreed with the research team when the research team actually judged
the label as correctly applied - i.e., the model rated that accuracy at
3 or lower but the team judged the label to be correct. We considered
one possible explanation to this phenomenon as the model was
conservative. Conversely, there were only three instances where the
model thought a label was accurately applied but the research team
believed the label was inaccurately labeled. Two of three instances
where the accuracy checking model believed the label model was correct
but the research team originally labeled them as incorrect were ``is
great!'' which received the label ``has good attitude'', and ``Hence, I
think that can perform better and help the team more by being more
outspoken about the topics being discussed'' which received the label
``made cognitive contributions''. This means that in general the model
tended to flag a small percentage of labels as potentially incorrect. In
practice, a user could then manually check this subset of flagged items
more than flagging items as correct when human raters might flag them as
incorrect (which happened less than 5\% of the time).

% Use the style of numbering in square brackets.
% If nothing is used, default style will be taken.
%\begin{enumerate}[a)]
%\item 
%\item 
%\item 
%\end{enumerate}  

%\begin{itemize}
%\item 
%\item 
%\item 
%\end{itemize}  

%\begin{description}
%\item[]
%\item[] 
%\item[] 
%\end{description}  

% Figure

%\begin{table}[<options>]
%\caption{}\label{tbl1}
%\begin{tabular*}{\tblwidth}{@{}LL@{}}
%\toprule
%  &  \\ % Table header row
%\midrule
% & \\
% & \\
% & \\
% & \\
%\bottomrule
%\end{tabular*}
%\end{table}

% Uncomment and use as the case may be
%\begin{theorem} 
%\end{theorem}

% Uncomment and use as the case may be
%\begin{lemma} 
%\end{lemma}


\hypertarget{discussion}{%
\section{Discussion}\label{discussion}}

Using a workflow in which we prompted the generative model to assign
topic labels to student comments suggested that the model could
accurately assign multiple topic labels in a single comment where
appropriate. The model was able to label 85\% of comments accurately by
identifying both the overt meaning as well as the underlying semantics
of the student comments. For example, comments mentioning
``communication'' and ``group meetings'' were labeled as such, but the
model also accurately labeled comments mentioning a teammate ``speaking
their mind'' or ``voicing issues'' as indicating communication and
cooperation. The ability of the model to make these connections across
varied word choices and statement structures suggests it has a nuanced
understanding of the themes that comprise student teammate feedback. It
also illustrates the utility of these models in contrast with rule-based
models that would require someone to create a dictionary of synonyms for
communication for a specific domain.

The most common inaccurate label, ``attended group meetings,'' may have
occurred because it was the first label option shown to the model,
leading the model to choose it by default when unsure. Negative comments
criticizing a lack of interaction or voicing issues were also
occasionally mislabeled as indicating cooperation, possibly because the
model focused on mentions of interaction and missed the negative
sentiment. To address these weaknesses, the model could be refined to
better account for sentiment and avoid defaulting to the first label
option. Another option could involve adding specific instructions about not defaulting to the first option if unsure. Alternatively, we, as researchers, could alter the label wording to specify that polarity. Doing so could help to pick up the valence of
the comment and lead to more accurate thematic labeling. We chose not to
do so this time to maintain fidelity with the original label taxonomies
from the literature.

These results compare favorably with other teammate feedback options
that involve using closed-ended options, where participants must choose
from a pre-determined list of labels. Scholars have proposed structured
prompts for students to reflect and respond to topics of specific
interests instead of freely open responses. For example, \citep{magana2022teamwork}
required students to write team retrospectives based on planning (e.g.,
what were team members' roles), monitoring (e.g., which aspects of the
team collaboration went well this semester), evaluation (e.g., what
do you think as a team was particularly good about the milestone you
just completed), and plan of action (e.g., what aspects do you think can
be done better for the next milestone regarding team performance). This
approach using a generative model to label student writing on open-ended
questions avoids constraining participants to the options provided by
researchers and allows for more nuanced, context-dependent labeling.
While not perfect, this approach does promise an opportunity for gaining
insights into participants' perspectives in their own words. The
benefits of a more generative approach to labeling student comment data
parallels the rationale that was behind the development of CATME's
behaviorally anchored system in the first place. More prescriptive
surveys of team behavior can be prohibitively long to address all the
relevant behaviors, resulting in survey fatigue, and yet team members
can still exhibit relevant behaviors that are not captured by surveys of
that kind. CATME's behaviorally anchored rating scale provides enough
detail for each dimension for team members to be able to classify other
behaviors as similar, without being overly prescriptive or causing
fatigue \citep{ohland2012comprehensive}.

On the note of improvement, continued iterative development could
further increase the accuracy and nuance of the model's labels.
Exploring ways to incorporate sentiment analysis and discourse context
into the model may help it better distinguish negative or ambiguous
comments. The accuracy-checking model could also be enhanced to provide
more specific feedback on incorrectly labeled comments to further
improve the labeling model.

Through this study, we provided evidence that pre-trained generative
text models show strong promise for thematically analyzing open-ended
comments when given a set of labels to use. This is important because
the model did not require any additional fine tuning yet still achieved
relatively high accuracy. When paired with the second step of rating its
own accuracy, we believe the model approaches the level of accuracy
needed for deployment in classroom settings. The model can identify
multiple, nuanced themes in single comments and across a spectrum of
language usage. With continued refinement, this approach could provide a
valuable tool for educational researchers to gain insights into
learners' experiences, perspectives, and interactions in online learning
environments. Meanwhile, researchers could also employ the usage of the
generative AI model in the process of data analysis as demonstrated in
this work.

\hypertarget{limitations}{%
\subsection{Limitations}\label{limitations}}

In this study, we focused on analyzing individual sentences from student
comments rather than full responses. While this approach allowed us to
evaluate the accuracy of ChatGPT in labeling comments for brevity, it is
possible that analyzing complete responses would provide additional
context and potentially improve the model\textquotesingle s performance.
Future research could examine the impact of considering full responses
and the interplay between sentences in a given response when identifying
topics in student comments.

Another limitation of our study is that we removed names from the
comments before analyzing them with ChatGPT. While this step was taken
to protect student privacy, names can sometimes provide valuable context
for understanding the nature of a comment. For example, a comment might
refer to a specific team member\textquotesingle s role or contribution,
and having the name in the comment could help the model more accurately
identify the topic. Future work could explore methods for anonymizing
names while preserving contextual information, which may further enhance
the model\textquotesingle s accuracy in analyzing student comments.

A third limitation is that the human raters each applied their own
assumptions and interpretations of the labels. For example, some judged
the classification results not only for the meanings associated with the
themes, but also the sentiment of the themes; some had different levels
of tolerance towards judging the ``closeness'' of the comments with the
ChatGPT classified themes. Therefore, from this experiment, we want to
emphasize the importance of generating clear, valid and reliable
classification models to use, for both human-only environments and
algorithms.

Furthermore, we used a predefined set of labels based on existing
literature that may have covered a large portion of the data. However,
there may be some aspects present in the data that are not captured by
the list of pre-existing labels, particularly new topics that evolve
with time and may not have been present when the labels were developed.
Future research can develop approaches which can identify cases in the
dataset which does align with any of the existing labels and can be
qualitatively examined by experts to develop new labels.

Finally, it is important to note that the current implementation of
ChatGPT is available only through API calls, making it inaccessible for
local use on one\textquotesingle s machine. There have been recent
releases of open source models such as Alpaca \citep{taori2023stanford}, Vicuna \citep{chiang2023vicuna},
and Koala \citep{geng2023koala}, but these have not achieved GPT-level performance
yet. This limitation of relying on an external service through an API
can pose challenges when working with sensitive data, such as student
comments containing personally identifiable information. It also is not
free and therefore has limited accessibility. As a result, using ChatGPT
for the analysis of student comments might not be feasible in all
educational settings. The development of local versions of the model, or
the implementation of privacy-preserving techniques, could help overcome
this limitation and broaden the potential applications of ChatGPT in
educational contexts.

\hypertarget{conclusion}{%
\section{Conclusion}\label{conclusion}}

Our study illustrates that pre-trained generative text models like
ChatGPT can effectively analyze open-ended student comments, achieving
85\% accuracy in assigning relevant topic labels. By adding a second
step in the workflow, where the model assesses its own accuracy,
instances of incorrect labeling can be further diminished, enhancing the
model\textquotesingle s value for deployment in educational settings.
While the current model is not perfect and necessitates additional
refinement, it holds significant potential in aiding instructors with
the efficient review and categorization of student feedback, enabling
them to concentrate on specific comments that demand their attention or
intervention.

It is crucial to emphasize that human judgment remains an essential
component of the process, as the model cannot entirely replace
instructors\textquotesingle{} input, particularly when addressing
complex or nuanced comments. Although there is room for improvement,
these initial results are promising regarding the potential utility of
this model as a tool for instructors in reviewing and interpreting
student feedback for more effective and tailored team management. The
model can help instructors swiftly sort through comments and pinpoint
those requiring closer attention or follow-up. With the inclusion of the
second step for accuracy checking, we believe that the need for
instructors to review every comment can be reduced to a small subset
flagged by the model as potentially inaccurate. Furthermore, refining
the model and training it on larger datasets could likely improve its
accuracy, as one might anticipate with future iterations of generative
models. Future research should investigate how accuracy might vary when
applying the model to diverse student populations, age groups, and
subject areas. In conclusion, this generative model can facilitate the
coding of student comments but cannot wholly replace human judgment.

\section{Acknowledgements}\label{acknowledgements}
This material is based upon work supported by the National Science Foundation under Grant No. 2107008.


\bibliographystyle{cas-model2-names}
\bibliography{catme-gpt-preprint}

\end{document}